\documentclass[a4paper,11pt]{article}
\usepackage{pos}
\usepackage{xspace}
\usepackage{wrapfig}
\usepackage{enumitem}
\usepackage{lineno}
\usepackage[capitalize]{cleveref}
\usepackage{siunitx}

\def\Offline{\mbox{$\overline{\textrm%
{Off}}$\hspace{.05em}\protect\raisebox{.4ex}%
{$\protect\underline{\textrm{line}}$}}\xspace}

\title{The core software and simulation activities for data analysis at the Pierre Auger Observatory}
\ShortTitle{Core software and simulation activities at the Pierre Auger Observatory}

\author*[a]{Eva Santos}
\onbehalf{for the Pierre Auger Collaboration$^b$}

\affiliation[a]{FZU - Institute of Physics of the Czech Academy of Sciences, Na Slovance 1999/2, Prague, Czech Republic}

\affiliation[b]{Observatorio Pierre Auger,
Av.\ San Mart{\'\i}n Norte 304, 5613 Malarg\"ue, Argentina \\
Full author list: \normalfont{\url{https://www.auger.org/archive/authors\_2024\_11.html}}}

\emailAdd{spokespersons@auger.org}

\abstract{The Pierre Auger Observatory, located near the town Malarg\"ue in the province of Mendoza, 
Argentina, is the largest cosmic-ray detector in existence, covering an area of \SI{3000}{\square\km}. 
The upgraded Observatory, in Phase II of operations, consists of a surface array of 1660
stations combining water Cherenkov, scintillator, and radio detectors. A subset of stations
also includes underground muon detectors. Additionally, fluorescence detectors located at 
four sites overlook the array. The science goals for the enhanced Observatory include the
measurement of the properties of ultra-high-energy cosmic rays with large statistics and
high sensitivity to the primary composition. The Observatory is also sensitive to photons
and neutrinos at the highest energies, allowing it to participate in multi-messenger
studies. The Auger \Offline Framework provides the tools to perform detailed simulations, 
using the Geant~4 toolkit, of all components of the Observatory and the analysis of both 
data and simulated events. It proved to have the flexibility needed to evolve during the 
lifetime of the Observatory, to accommodate new sub-detectors and, recently, changes to the 
station readout electronics. A new challenge is interfacing the framework with Machine 
Learning tools for both the development and execution of neural-network-based algorithms. 
Independent of the framework, CORSIKA~7 is used to simulate particles, fluorescence light, 
and radio signals produced by air showers. The production of simulations is coordinated 
centrally to provide standard libraries for analyses and to optimize the use of computing 
resources. We will describe the evolution and status of the \Offline Framework and the tools 
used to coordinate the simulation efforts. We will also discuss the challenges of the 
massive simulation efforts and the resources consumed to provide the simulation libraries 
required by the Collaboration.}

\FullConference{7th International Symposium on Ultra High Energy Cosmic Rays (UHECR2024)\\
 17-21 November 2024\\
Malargüe, Mendoza, Argentina\\}

\begin{document}
\maketitle

\section{Introduction}
The Auger \Offline Framework~\cite{Argiro:2007qg} is one of the keystones of the core software used by the Pierre Auger Collaboration. 
Its origin dates back to 2002, and, already back then, the Auger \Offline was aimed at building software sturdy enough to endure more 
than 20 years, comprising the whole life of the Pierre Auger Observatory, in which further upgrades would be a reality. 
The software framework consists of an object-oriented C++-based code designed to be flexible and robust enough to support the simulation 
and reconstruction of all types of events detected by the Observatory~\cite{PierreAuger:2015eyc}. 
It can handle several types of data formats and databases, encompassing event simulation and reconstruction, as well as monitoring information 
from all the instruments. 
For the event simulation, the reading of the output files from several extensive air shower Monte Carlo codes, namely CORSIKA~\cite{Heck_corsika:a} 
and AIRES~\cite{Sciutto:1999jh}, with the options CoREAS~\cite{Huege:2013vt} and ZHAiRES~\cite{Alvarez-Muniz:2011ref} for the radio component of 
air showers, and also CONEX~\cite{Pierog:2004re}, for fast simulations of the longitudinal development of air showers, is supported. 
The advent of AugerPrime~\cite{PierreAuger:2016qzd}, the major upgrade of the Pierre Auger Observatory, was the biggest challenge overcome 
by the Auger \Offline Framework, as the implementation of some of its components was found to break the original framework 
philosophy~\cite{PierreAuger:2021fah}. 
Most of the production of the extensive simulation libraries used by a large number of analyses in the Collaboration is done using the grid 
through the Virtual Organization Auger~\cite{PierreAuger:2023cqe}.


\section{Core software}
The core software of the Pierre Auger Collaboration consists of a set of programs dedicated to the study of high- and ultra-high-energy 
cosmic rays. 
At the head is the Auger \Offline Framework~\cite{Argiro:2007qg}, a modular and universal software used for the reconstruction and 
simulation of the events detected by the Pierre Auger Observatory, which we detail below.
The remaining core software employed by the Pierre Auger Collaboration consists of the Monte Carlo programs CORSIKA~\cite{Heck_corsika:a},
AIRES~\cite{Sciutto:1999jh}, and CONEX~\cite{Pierog:2004re}, which are used to simulate the development of extensive air showers and whose 
output files are given as input to the Auger \Offline Framework for the event simulation.
Other programs used are CRPropa~\cite{AlvesBatista:2022vem} and SimProp~\cite{Aloisio:2017iyh}, aimed to simulate the propagation of 
cosmic rays in an extragalactic environment.

\subsection{Auger \Offline Framework}
The Auger \Offline is a universal framework for the reconstruction and simulation of events detected at the Pierre Auger Observatory. 
It is an object-oriented C++ code with a modular structure aimed at being flexible and robust software, allowing it to be easily extensible 
to accommodate upgrades to the Pierre Auger Observatory instrumentation throughout its entire lifetime. 
It was also meant to be a collaborative effort of a large number of physicists developing a variety of applications regarding the 
reconstruction and simulation of all types of events detected at the Pierre Auger Observatory. 
For this purpose, the parts of the code directly used by physicists should be clear and relatively straightforward to modify.

\subsubsection{The upgraded framework} 
The initial design of the Pierre Auger Observatory envisaged a hybrid detection technique comprising a \SI{3000}{\square\km} 
\emph{surface detector array} (SD), composed of $1600$ \emph{water-Cherenkov detectors} (WCD) arranged in a triangular grid of 
\SI{1500}{m} spacing (SD-1500), observed by the \emph{fluorescence detector} (FD), consisting of $24$ telescopes placed in four 
locations on the SD periphery. 
The first enhancement to the Pierre Auger Observatory targeted an extension to lower energies by deploying $61$ additional WCDs 
in a \SI{23.5}{\square\km} denser array of \SI{750}{m} spacing (SD-750) and HEAT - \emph{High Elevation Auger Telescopes}, three 
fluorescence telescopes at the Coihueco site, extending the FD elevation range from $30^{\circ}$ to $58^{\circ}$. 
The \emph{Auger Engineering Radio Array} (AERA) was later added to the SD array, serving as proof of concept of the feasibility 
of the future radio detector (RD) of AugerPrime, which incorporates a \emph{short aperiodic loaded loop antenna} (SALLA) from the RD, 
and the \emph{surface scintillator detector} (SSD) atop most WCD. 
Moreover, the dynamic range of the SD was extended by the installation of a small photomultiplier tube inside each station, and a 
new electronics board with faster sampling connecting all the detectors was added. 
Finally, the \emph{underground muon detector} (UMD), an array of \SI{30}{\square\m} area scintillator detectors buried at a 
depth of \SI{2.3}{m} at the SD-750 and the newer SD-433 array, in the vicinity of a WCD, aimed at the direct measurement of the 
muon content of air showers~\cite{PierreAuger:2016crs, PierreAuger:2017aph} was deployed. 
While some changes were straightforward to implement, the implementation of AERA required significant additions to the 
software~\cite{PierreAuger:2011btp}. 
Also, some of the AugerPrime components posed challenges to the framework as it meant breaking with some old paradigms of 
the Auger \Offline, called for some unexpected changes, and an update from C++ 98/03 to C++ 11/14 to break strict 
backward compatibility, eliminate deprecated interfaces, and modernize the development 
infrastructure~\cite{PierreAuger:2021fah, PierreAuger:2023cqe}.\\
Currently, the Auger \Offline can be used to reconstruct events taken during the old (Phase I) and the \mbox{AugerPrime}
(Phase II) configurations. 
The reconstruction and simulation of events comprises those only detected by the SD (SD-1500, SD-750, and SD-433), 
the FD, and hybrid events, i.e., those detected simultaneously by the SD and the FD, including the HECo extension 
(HEAT and Coihueco telescopes), and events detected by the AERA, the first radio extension to the Observatory. 
For Phase II, the novelty is the possibility of the reconstruction of multi-hybrid events, i.e., events detected by 
the WCD from the SD, including the SSD, the RD, and the UMD.

\subsubsection{Structure}
The Auger \Offline package is organized to have a strict acyclic dependency to avoid problems when building the code. 
Its structure comprises three main parts, namely, the Modules, the Event, and the Detector, described in detail below. 
These three components are complemented by a set of foundational classes and utilities for error logging, physics, and mathematical 
manipulations, as well as packages that support abstract manipulation of geometrical objects. 
The latter ones do not require any knowledge of the Pierre Auger Observatory or cosmic-ray physics. 

\begin{figure}
    \centering
    \vspace{-0.3cm}
    \includegraphics[width=.4\textwidth]{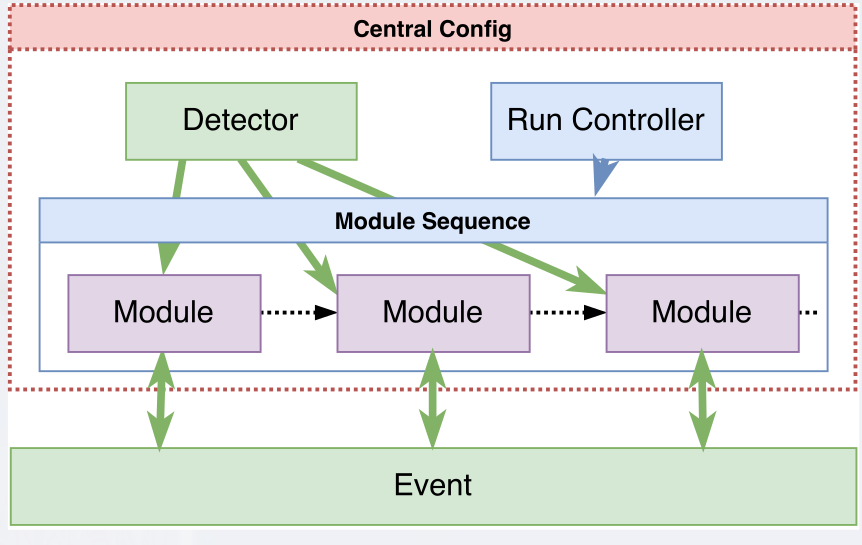}
    \caption{Schematic sketch of the \text{Central Config} of the framework defining the Module configuration, sequencing, and the Detector.
    The \texttt{Run Controller} defines the sequence in which the Modules are executed.}
    \label{fig:1}
\end{figure}    

\paragraph{Modules}The Modules can be assembled and sequenced using instructions in XML format. 
At the beginning of a run, the framework processes a hierarchy of XML files, starting from the bootstrap file that links all the other 
necessary XML files for the configuration of the Detector and establishes the module sequence to be executed, see~\cref{fig:1}.
This XML hierarchy is also used to provide additional configuration data required by the individual modules. 
The reading and writing operations are handled by special modules, which provide a thin interface to the event input/output (I/O) layer. 
The names of the files to be processed and written into are part of the data given to the modules.

\begin{figure}
    \centering
    \includegraphics[width=.8\textwidth]{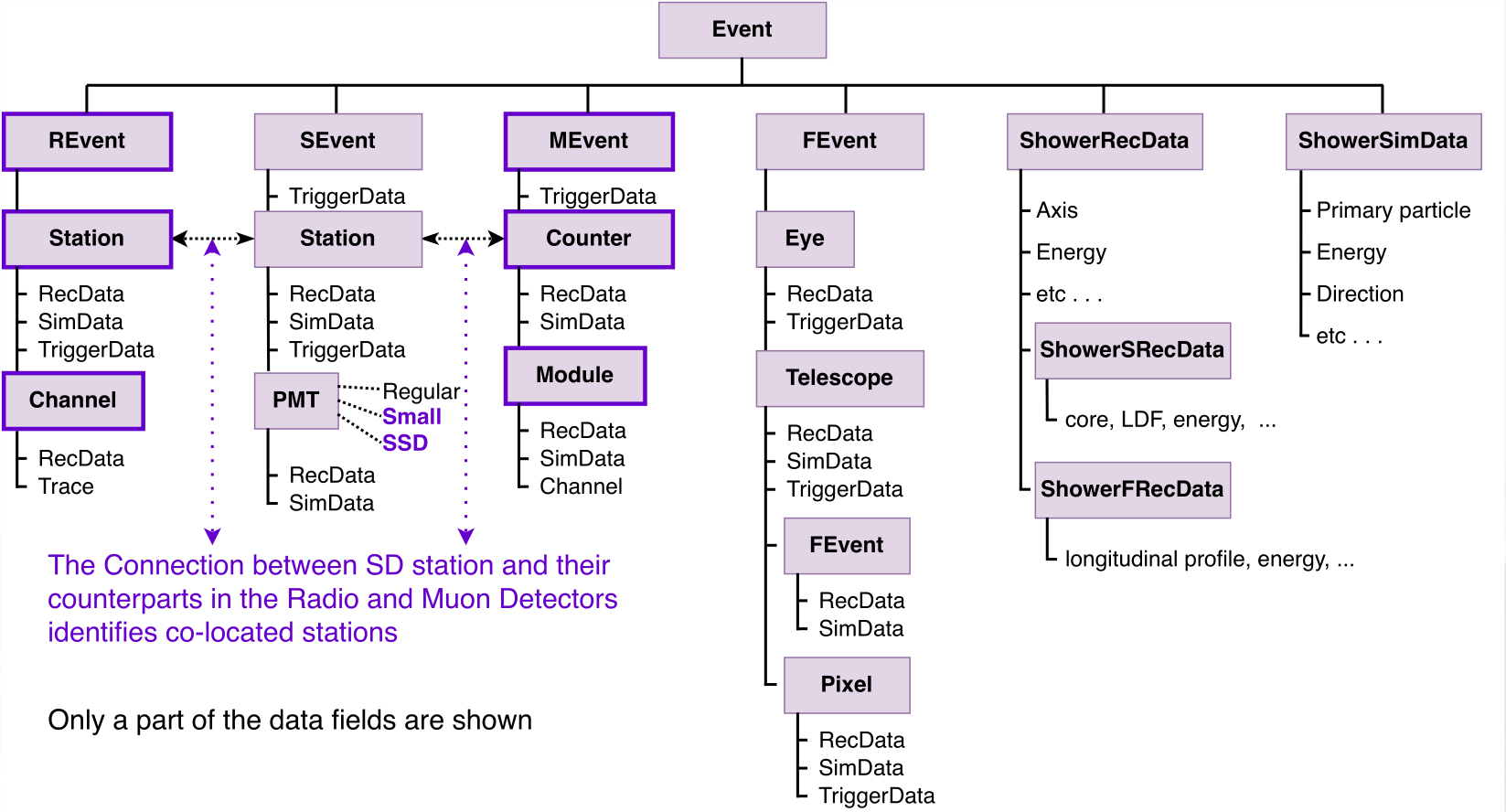}
    \caption{Schematic sketch of the Event data structure. The AugerPrime additions are given in purple. Not all components are shown.}
    \label{fig:2}
\end{figure}

\paragraph{Event}The Event serves as a central data structure where modules store and retrieve information, accumulating all reconstruction 
and, when applicable, also simulation information. 
It is organized as a collection of classes that mirror the instrument layout of the Pierre Auger Observatory, as well as further subdivisions 
for accessing Monte Carlo parameters. 
The Event structure contains all raw, calibrated, reconstructed, and Monte Carlo information, acting as the core framework for communication 
between modules, as illustrated in~\cref{fig:2}. 
The Event data structure has to be filled with the data expected by the module sequence at the beginning of the Event loop, typically by 
reading information from one of several available sources. 
For data events, these formats include the raw event, which is the data format from the DAQ, the internal \Offline format, and monitoring 
formats, where there is information about the raw and calibrated data, as well as the results from the different stages of the event reconstruction. 
For simulations, the reading of output files from the Monte Carlo codes CORSIKA~\cite{Heck_corsika:a}, AIRES~\cite{Sciutto:1999jh}, 
and CONEX~\cite{Pierog:2004re} is supported. 
In this case, the Event structure also includes information related to detector simulations. 
Most detailed detector simulations make use of the Geant4 package~\cite{Ivanchenko:2003xp, Allison:2006ve, Allison:2016lfl} complemented by 
custom simulations for the tracking of the Cherenkov light, fluorescence photons, and the simulation of the electronics. 
The Event serialization is implemented using the ROOT toolkit~\cite{Brun:1997pa}. 
The internal ROOT-based format allows the user to save and restore the complete information in the Event data structure. 
Similarly, it is also possible to save the result of simulations in the raw data format used for data acquisition or in the 
\emph{Advanced-Data Summary Tree} (ADST) format~\cite{Gonzalez:2012vj}. 
The latter is used for the front-end user analysis based on the I/O component of the ROOT framework, and it is a lightweight 
tool that can be used in the later stages of data analysis.

\begin{figure}
    \centering
    \includegraphics[width=.8\textwidth]{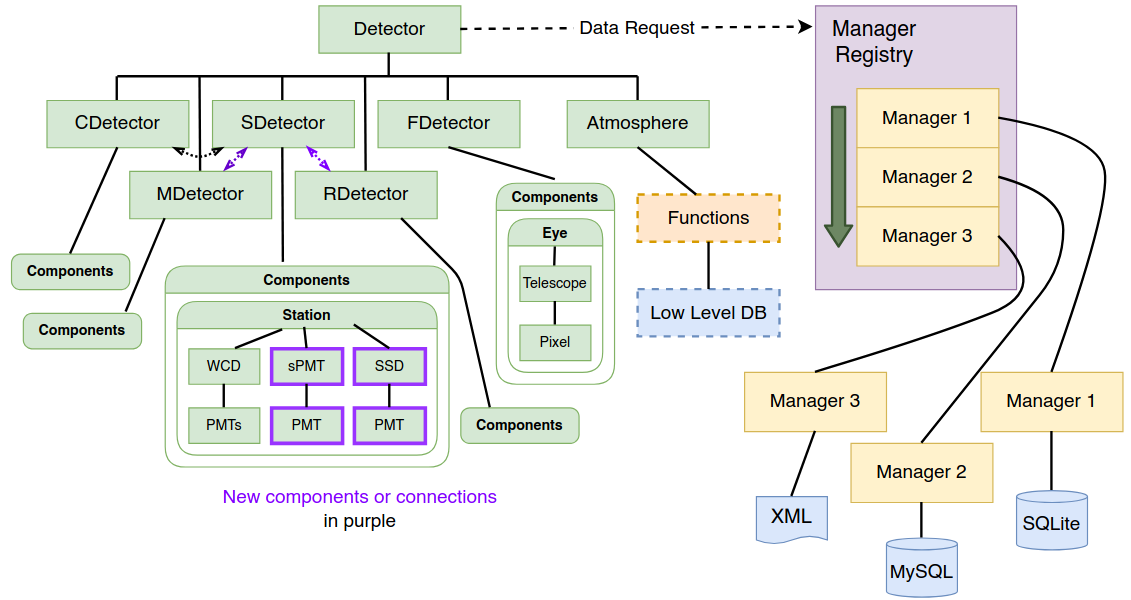}
    \caption{Schematic sketch of the Detector structure, including the access to the atmospheric data. The AugerPrime additions are shown 
    inside the purple boxes.}
    \label{fig:3}
\end{figure}

\paragraph{Detector}The Detector description provides a gateway to data, portraying the detector configuration and performance. 
Since the atmospheric conditions at the time of the event detection impact the development and properties of extensive air showers, data 
from the atmospheric monitoring as a function of time are also included as part of the Detector. 
The Detector components are mapped directly onto data structures. 
Similarly to Event, the Detector interface follows the hierarchy associated with the instruments of the Pierre Auger Observatory, as 
shown in~\cref{fig:3}.
A set of easily usable functions to extract the data is provided. 
The managers provide access to different data sources; in particular, static detector information is stored in XML format, while 
time-varying monitoring and calibration data is stored in MySQL and SQLite databases. 
The Detector structure acts as a facade providing user-visible interfaces to configurable managers, enabling flexible access to various 
data sources.

\subsubsection{Next steps}
The increasingly frequent use of \emph{machine learning} (ML) algorithms in a wide range of applications is 
transforming the traditional way we do scientific analysis, a phenomenon to which the Pierre Auger Collaboration is no 
exception.
See, for instance,~\cite{PierreAuger:2024flk, PierreAuger:2024nzw, hahn:uhecr2024, rodriguez:uhecr2024}. 
In this context, we are implementing a connection between the \emph{Open Neural Network Exchange}~\cite{onnx} (ONNX) ecosystem.
This open-source framework provides a standardized representation for ML models and the Auger \Offline.
Other plans include integrating simulations with the CORSIKA 8~\cite{c8:uhecr2024} program into the framework.


\section{Virtual Organization Auger}
The Virtual Organization (VO) Auger, established in 2006, is a group of institutions sharing a common goal and collaborating using 
distributed computing and data resources provided by the European Grid Infrasture (EGI). 
The central resources, such as the registration portal and the VOMS (\emph{Virtual Organization Membership Service}) server, are 
maintained by the CESNET MetaCentrum. 
All members of the Pierre Auger Collaboration can apply for membership by filling out a registration form, which must be approved by the 
VO manager. 
The membership has a validity of approximately one year, after which it can be renewed upon request.\\
Since 2014, the VO Auger has used the DIRAC (Distributed Infrastructure with Remote Agent Control) interware for job submission, job 
monitoring, and file catalog management. 
Our DIRAC server is lodged at the Frances Grilles Infrastructure.

\subsection{Grid usage: October 2023 - October 2024}
\begin{wrapfigure}{r}{0.5\textwidth}
    \centering
    \vspace{-0.8cm}
    \includegraphics[width=.275\textwidth]{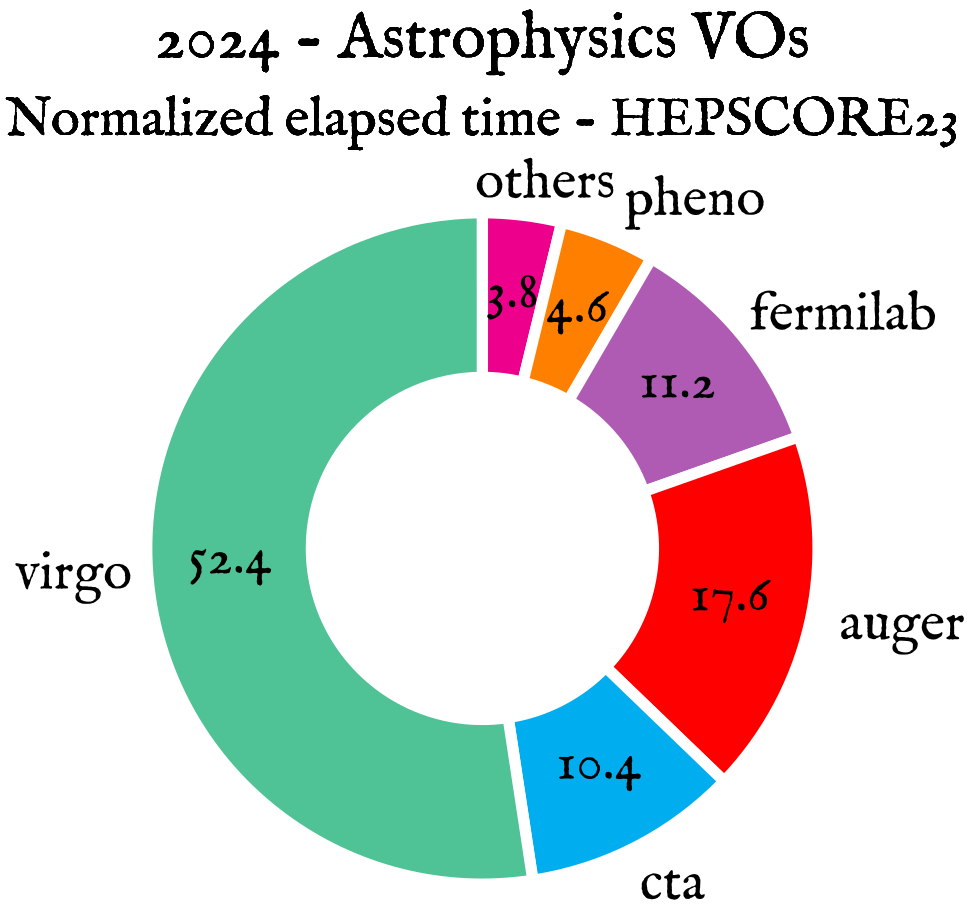}
    \caption{Relative elapsed time of the Astrophysics VOs in 2024, according to the EGI accounting portal. 
    The contribution of the VO LHCb was excluded.}
    \label{fig:4}
\end{wrapfigure}
\begin{figure}[h]
    \centering
    \includegraphics[width=.45\textwidth]{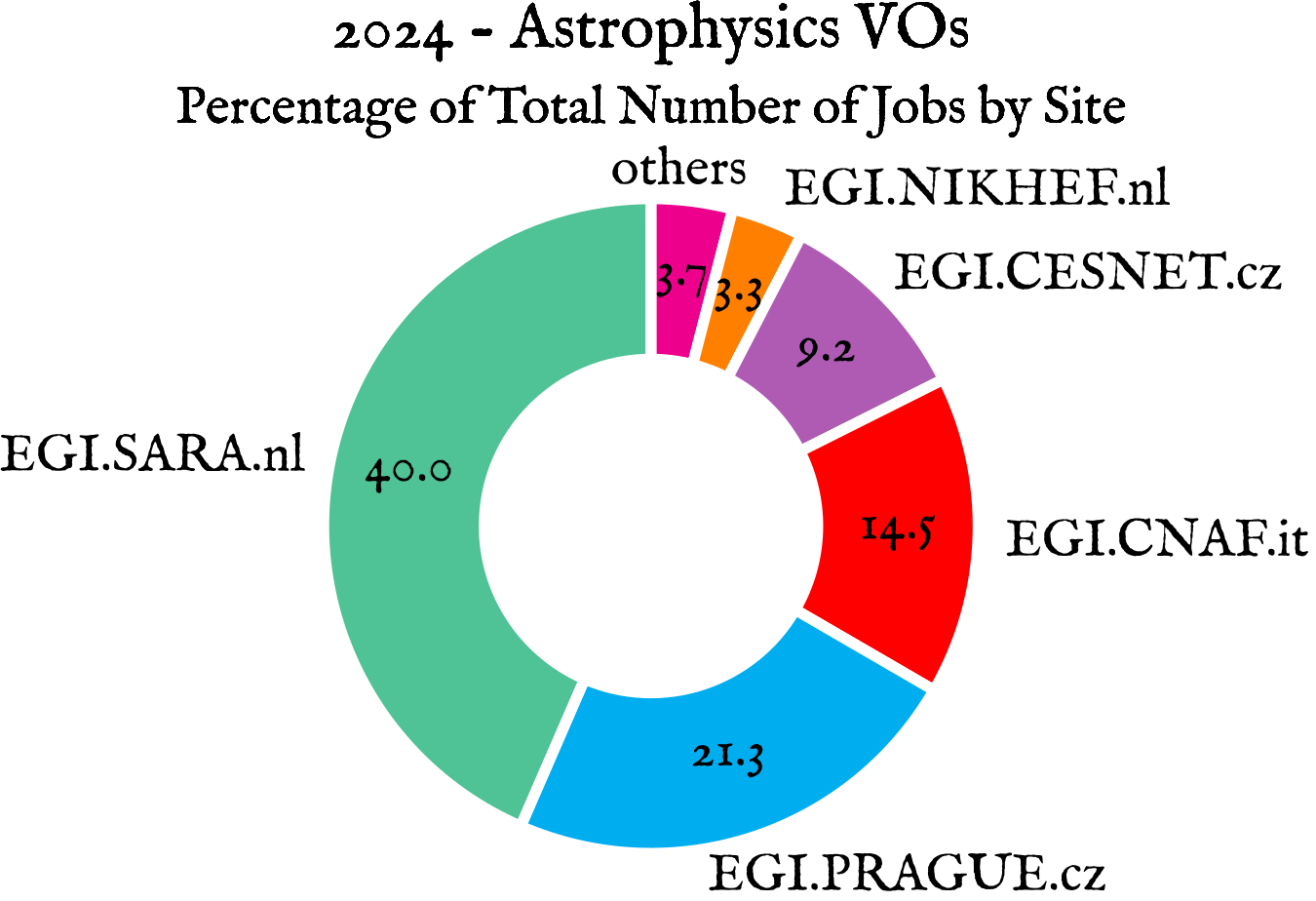} \hspace{1.0cm}
    \includegraphics[width=.45\textwidth]{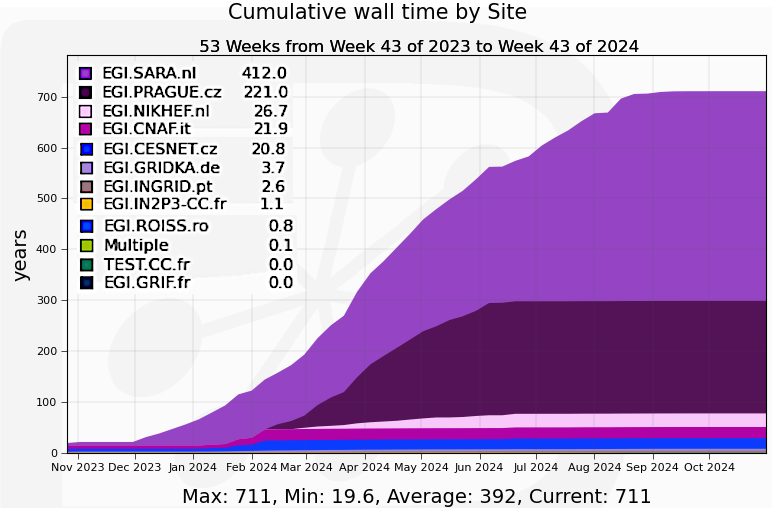}    
    \caption{Left: Relative distribution of jobs per grid site. Right: Cumulative wall time per site.}
    \label{fig:5}
\end{figure}

According to the \href{https://accounting.egi.eu/discipline/Astrophysics/normelap_processors/VO/Year/2023/10/2024/10/}{EGI accounting portal}, in 2024, the 
VO Auger ran about 855 thousand single-core jobs on 10 grid sites from 7 countries, totaling 178 million CPU hours, normalized to HEP-Score23 
(HS23)~\cite{Sobie:2023ftp}. 
Excluding the VO LHCb from the list of Astrophysics VOs, the VO Auger appears as the second largest EGI user with a relative usage of 17.6\%, only behind 
VIRGO, with 52.4\%, as shown in~\cref{fig:4}. 
The VO Auger is consolidating its position as one of the largest users, having increased its HS23 from 77 to 178 million normalized CPU hours in the 
2020 - 2024 period. 
In~\cref{fig:5}, the distribution of the relative number of jobs per grid site (left) and the cumulative wall time by site (right) are shown. 
The bulk of the grid usage in this period comprises the production of CoREAS~\cite{Huege:2013vt} simulations. 
The EGI.SARA.nl site led in both the number of jobs executed and cumulative wall time. 
Two main factors explain this: only this site accepted long jobs, and possibly the end-of-time of CentOS7, the operating system used to compile 
CORSIKA and CoREAS considerably worsened the performance of these jobs. 
From the experience learned, we advocate for containerizing all software in the CVMFS - \emph{Cern Virtual Machine File System} repository to enhance 
job performance, particularly in a potentially heterogeneous distributed computing environment.


\section{Monte Carlo simulation libraries}
The Pierre Auger Collaboration has a task dedicated to the production of the shower and detector official simulation libraries used in a wide variety of 
physics analyses. 
Currently, air shower simulations are produced with the CORSIKA~\cite{Heck_corsika:a} program, whose output files are fed into the Auger \Offline 
Framework~\cite{Argiro:2007qg}. 
So far, the bulk of our detector simulations concern Phase I of the Observatory and comprise SD-only for the SD-1500, SD-750, SD-433, and also 
hybrid simulations, namely, FD together with SD-1500 and HECo with the SD-750. 
Special detector simulation libraries comprise hybrid and SD-only time-dependent simulations, which try to reproduce the performance of the SD and FD 
detectors over time. 
A detailed description of our shower simulations can be found in~\cite{PierreAuger:2023cqe}. 
Since then, according to recent findings~\cite{Zhang:uhecr2024}, the cosmic-ray shower library was extended to include Tellurium, Platinum, and Uranium-induced 
air showers in the energy range $10^{17}$ - $10^{20.2}\,\mathrm{eV}$, by using the option CONEX in CORSIKA for the hadronic interaction model 
EPOS-LHC~\cite{Werner:2005jf, Pierog_2015}. 
Uranium-induced air showers are meant to train ML algorithms on the muon content~\cite{hahn:uhecr2024}. 
Besides the production of the official libraries, the task is also open to simulation requests for specific studies, such as the case of the work reported 
in~\cite{Ellwanger:uhecr2024}, or in aiding Collaborators with limited computing resources, as long as their requests are well justified and do not clash 
with the priorities of the task. 
The bulk of the simulations are produced using the grid resources provided by the VO Auger. 
However, for a fraction of CoREAS, whose required CPU time exceeds one week, and other non-standard or test productions, the Prague cluster, which has a 
maximum wall time of 30 days, is used.

\subsection{Accessing the simulations}
The preferred method for accessing simulations has been the iRODS (\emph{integrated Rule-Oriented Data System}) provided by the Computing Center 
IN2P3 in Lyon, France, which will cease its support for the Collaboration at the end of 2025, being replaced by the CNAF Computing Center 
in Bologna, Italy. 
Meanwhile, some collaborators have started using the DIRAC Data Management System (DMS), provided they have a valid X509 certificate. 
DIRAC offers faster file transfers compared to iRODS, as older files at CC IN2P3 are stored on tape, which may significantly increase the download 
time of each file. 
However, the issuing of X509 certificates is not possible for all the Collaborators. 
This problem could be overcome if DIRAC replaces the certification process, such as by issuing tokens for login or making the simulations available 
in CNAF. 
Currently, there are about 11 million files with a total size of \SI{1}{PB} registered in the DIRAC File Catalog.

\subsection{Next steps}
For 2025, we plan to redo the whole cosmic-ray shower library with the newest CORSIKA 7.7600, which will contain updated 
hadronic interaction models. 
The low-energy cut for hadrons will be reduced from $0.05$ to \SI{0.02}{\GeV} as it proved to reproduce the late pulses 
observed by the SSD better~\cite{Schulz:uhecr2024}. 
This lower energy cut is enabled by the FLUKA~\cite{Ferrari:2005zk, Ballarini:2024uxz} versions 2024.1.0 and above. 
The number of proton-induced showers will be increased from 5000 to 10000 as required for signal-to-background studies for 
the searches of neutral particles, such as photons and neutrinos. 
We also plan to produce a tau-neutrino-induced air shower library, with and without the radio component of air showers. 
For long CoREAS simulations, we will exploit the possibility of running MPI jobs on the grid. 
Finally, we will update our Phase I and Phase II detector simulations using the Auger \Offline.

\section*{Acknowledgements}
 This work is co-funded by the Czech Science Foundation under the project 
 \mbox{GA\u{C}R~24-13049S}, and by the European Union and by the Czech Ministry of
Education, Youth and Sports - Project No. FORTE – CZ.02.01.01/00/22\_008/0004632.

\end{document}